# Relative Lempel-Ziv Factorization for Efficient Storage and Retrieval of Web Collections

Christopher Hoobin[1], Simon J. Puglisi[1,2], and Justin Zobel[3]

[1]School of Computer Science and Information Technology, RMIT University
[2]Department of Informatics, King's College London
[3]Department of Computer Science and Software Engineering, University of Melbourne

christopher.hoobin@rmit.edu.au, simon.puglisi@kcl.ac.uk, jzobel@unimelb.edu.au

## ABSTRACT

Compression techniques that support fast random access are a core component of any information system. Current state-of-the-art methods group documents into fixed-sized blocks and compress each block with a general-purpose adaptive algorithm such as GZIP. Random access to a specific document then requires decompression of a block. The choice of block size is critical: it trades between compression effectiveness and document retrieval times. In this paper we present a scalable compression method for large document collections that allows fast random access. We build a representative sample of the collection and use it as a dictionary in a LZ77-like encoding of the rest of the collection, relative to the dictionary. We demonstrate on large collections, that using a dictionary as small as 0.1% of the collection size, our algorithm is dramatically faster than previous methods, and in general gives much better compression.

## 1. INTRODUCTION

Storage of collections of data is one of the most challenging problems of the information age [3]. Here, compression plays a central role and is a fundamental component of any information retrieval system [9, 14, 23, 30, 34]. Compression improves both search and retrieval by reducing the effect of disk-seek time and read latency, increasing bandwidth between levels of the memory hierarchy [8, 26, 35].

In the context of text retrieval, a compression algorithm first must maintain a compact representation of the collection. Second, it must provide fast random access to specific documents of the collection for retrieval and post-processing tasks, such as query-biased snippet generation [27, 28, 29]. Generally decompression time is far more important than compression time as usually a document will be encoded once but decoded many times. However, the compression algorithm must be practical and scalable.

A standard existing approach is to store documents in blocks, and then compress the blocks with a library, such as that provided by ZLIB. This approach implies a classical trade-off between speed and time. With small blocks, there is insufficient data for the compressor to learn a representative dictionary, and thus compression is poor. With a large block, on average half of which must be decompressed to access an individual document, retrieval speed is compromised.

We propose a novel, yet straightforward solution. Standard LZ methods use the data to be compressed as a dictionary. In our method, we generate a dictionary by sampling the whole collection, and then encode each document by reference to this dictionary with an LZ-like coding. Dictionary size is a parameter, but, as we show can be usefully held to a small fraction of current laptop memory sizes. Our results show that our approach achieves decompression and random access at speeds dramatically faster than previous alternatives. For data that can be sorted in a way such that similar documents are adjacent, for example by URL sorting [16], existing block-oriented methods can yield better compression at the cost of extremely slow retrieval; otherwise our method also provides superior compression. This is primarily due to the fact that our method can exploit non-local redundancy.

The paper is organized as follows: The next section reviews related work. In Section 3, we provide details of our compression scheme (dictionary generation, encoding and decoding algorithms). We report on an experimental evaluation of our approach in Section 4. Section 5 provides a discussion of our results. Section 6 offers conclusions and outlines future work.

## 2. BACKGROUND

Our aim is to develop compression techniques that are effective for corpora of text whose sizes mean that they must be stored on disk. In this section we review semi-static and adaptive compression methods for large collections of this kind. Our specific focus is on methods that provide fast decompression and allow for reasonably efficient random access to arbitrary documents in the compressed collection.

### 2.1 Semi-static Methods

A semi-static compressor makes two passes over the collection. The first pass is to calculate the frequency of symbols in the text, such as characters, word–nonword pairs, or phrases. These statistics are used in the second pass to assign codewords to each symbol, with frequent symbols assigned shorter codewords and rarer symbols assigned longer codewords. Traditionally, codewords were assigned to symbols at a character level by a bit-oriented Huffman code [18].

This generally leads to poor compression of natural language texts to around 60% of their original size [25]. Using words as symbols leads to much better compression, as the word distribution of natural language texts is much more biased than the character distribution [34]. A word-based Huffman code can compress a natural language text to nearly 25% [24, 30]. However, we note that the supporting experiments were undertaken on data without significant mark-up and other non-text content, which can drastically increase the number of 'non-words' and thus the memory requirements for the dictionary.

De Moura et al. [15] describe two byte-oriented coding techniques, Plain Huffman (PH) and Tagged Huffman (TH), that provide much faster encoding and decoding speeds at a cost of slightly reduced compression performance (5% to 10%) compared to bit-oriented codes. Operating at a byte level eliminates the need for expensive bit manipulations required in traditional Huffman coding. In Tagged Huffman codes a flag bit reserved in each byte is used to signal the start of a codeword. This allows for fast compressed pattern matching. A pattern can be encoded with the same model and searched for directly in the compressed text.

Brisaboa et al. [5] propose their End Tagged Dense Code (ETDC), an improvement to Tagged Huffman coding, by modifying the flag bit to symbolize the end of a codeword. This reduces the requirement to build a Huffman tree to ensure each symbol is a valid prefix code. Dense Codes provide an acceptable space trade-off to Tagged Huffman codes. They are simpler to implement and are faster at compression and decompression. Like the other Huffman approaches, it is necessary to build an explicit dictionary of symbols.

There is a family of dense codes described in literature that can be used for these kinds of compression tasks. Brisaboa et al. [4] describe Pair-Based and Phrase-Based End-Tagged Dense Codes (PETDC and PhETDC). These are two extensions to ETDC that use symbols of a higher order. In PETDC symbols can be either words or pairs of words. In PhETDC symbols are considered words or phrases of words. It is reported that PETDC can reduce a text by 70%, and PhETDC by 77%, outperforming all current zero-order word-based semi-static compressors [4]. A Dynamic End Tagged Dense Code (DETDC) [6] is a dynamic version of ETDC, where the model is transmitted along with the encoding, much like an adaptive compression algorithm. A Dynamic Lightweight End Tagged Dense Code (DLETDC) [7] is a modification to this scheme that reduces the cost of transmitting the model.

Compressing large collections with a semi-static model can be problematic, as the vocabulary can be significantly larger than physical memory. Our parsing of Clueweb09 Category A (a 15TB English text web crawl) generated a 13GB uncompressed vocabulary. Close to 50% of the lexicon was comprised of non-word symbols that occurred only once throughout the collection. Such hurdles to scalability have not been reported before in literature due to the relatively small collections sizes used in experiments. For example, Turpin et al. [29] use a semi-static model for document compression, but the collections they used were less than 100GB.

In general, overall compression achieved by a semi-static approach is limited by its inability to take advantage of any global repetitive properties of the collection. The best case for these methods is a reduction of the text size to at least 20% of the original, which is rather worse than the 9%–14% we report for our methods on much smaller dictionaries.

## 2.2 Adaptive Methods

Popular adaptive methods for document compression use dictionary algorithms based on the Lempel-Ziv (LZ) family [31, 32]. This family forms the foundation of many popular compressors, including DEFLATE, ZLIB, LZMA, and PNG. These tools exploit local duplication in a text through the use of a small sliding window which acts as a dictionary. Encoding involves replacing substrings with pointers to previous occurrences found in the dictionary. There are many methods for selecting matches in a dictionary. When the dictionary is small (for example ZLIB typically uses a dictionary of less than 32KB), the most effective method is to hash all possible strings and to select the longest match. As the dictionary fits comfortably in higher levels of cache (L2 cache sizes on a current CPU range in megabytes), using more advanced methods would incur additional costs. When using a larger window it is better to create an index over the dictionary, such as a suffix array [22], to improve selection time.

Encoders using a small window achieve only moderate compression on larger collections as they are not taking advantage of any global repetitive properties across the collection, such as boilerplate markup in web crawls or highly repetitive genetic databases. For there to be an improvement over existing LZ algorithms we need to capture global repetition. Extending the dictionary will lead to an obvious improvement in compression. Ferragina and Manzini [16] demonstrate that applying LZMA with a dictionary size of 128MB compresses GOV2, a 426GB web crawl, to as little as 4.85% of its original size. However, increasing the dictionary size hinders random access, as dictionary decoding must always start from the beginning of a block.

Open source information retrieval systems such as Lucene[1] and Indri[2] use a blocked approach for document storage and retrieval. Collections are split into fixed size blocks and compressed with an adaptive algorithm (ZLIB). This introduces a trade-off between block size and decoding time. Storing a single document in a block will lead to faster document retrieval speeds at the cost of compression. Grouping documents into blocks will lead to better compression, as there is greater redundancy to exploit, but reduces retrieval speeds, as on average more documents must be decompressed on the way to retrieving a requested document.

Note that document storage and fast document retrieval are not the primary goal of either IR system. However, this method is a practical approach. Google employs a similar method, which they outline in a paper describing their distributed storage system Bigtable [12]. First, all pages are clustered together according to their host name and then compressed in two passes. In a initial pass they pre-process document clusters with Bentley and McIlroy's algorithm described in [2]. This algorithm is especially effective and can eliminate a lot of redundancy on sorted collections. In a second pass they use a "fast compression algorithm that looks for repetitions in a small 16KB window", which we assume is something akin to ZLIB (details are not provided).

Kreft and Navarro [19] recently described LZ-END, an LZ77-like algorithm that directly supports random access. This approach is not viable for large document collections as a suffix array and other auxiliary data structures of the complete collection are required for encoding. Another set of related techniques are grammar compressors, such as RAY [10], XRAY [11] and RE-PAIR [21]. Grammar compressors can achieve powerful compression but have enormous construction requirements, limiting their application to smaller collections.

## 3. RELATIVE LEMPEL-ZIV FACTORIZATION

At the core of our method is a string data structure we call the *relative Lempel-Ziv factorization* [33, 20]. Let $x = x[1..n]$ be a string of length $n$, and $d = d[1..m]$ be a string of length $m$, where $m \leq n$. The relative LZ factorization $RLZ_x$ of $x$ is a factorization of $x$ into substrings, $x = w_1 w_2 \ldots w_k$, relative to $d$; such that each substring $w_j$, $j \in 1 \ldots k$, is either:

1. The longest factor (ie. substring) of $d$ starting at the current position in $x$, $|w_1 w_2 \ldots w_{j-1}|$; or

2. a single character $c$ from $x$ that does not occur $d$.

Each factor $w_j$ is represented as a pair $(p_j, l_j)$, where $p_j$ specifies an offset to a position in $d$ and $l_j$ denotes the length of the factor in $d$. If $l_j = 0$, $p_j$ contains a character $c$ that does not occur in $d$.

As an example, the LZ factorization of the string $x = bbaancabb$ relative to a dictionary $d = cabbaabba$ from Table 1 will compute three pairs: $(3, 4)$, corresponding the string *bbaa* at offset 3 and length 4 in $d$, ('n', 0), as the character $n$ does not exist in the dictionary, and $(1, 4)$, the string *cabb* beginning at offset 1 in $d$.

### 3.1 General Approach

Our compression scheme, RLZ, operates as follows.

1. Construct a dictionary, $d$, by concatenating a (random) sample of documents from the collection, of total length $m$ bytes. Here, $m$ is dictated by the user and/or the available memory.

2. For each document, $t$, in the collection, factorize $t$ relative to $d$ into $(p_i, l_i)$ factors (or pairs) and encode each pair efficiently (details in Section 3.4).

3. Store a document map which provides the position on disk of each encoded file. This component is common to all large scale file compression systems.

4. Access to a desired document is achieved by first locating the starting position of the document with the document map, and then decoding the $(p, l)$ pairs, translating each into text via the dictionary, $d$.

RLZ provides fast random access to documents because the dictionary is no longer adaptive, and can be made small enough to be held resident in memory. For effective compression the dictionary must capture the overall structure of the collection, that is, global repetitions. We describe a simple yet highly effective dictionary generation technique in Section 3.3.

### 3.2 Scalable Compression

We can compute the RLZ factorization in $O(n \log m)$ time and $O(m)$ words of memory, using an algorithm inspired by the CSP2 algorithm for LZ77 factorization by Chen et al. [13]. The main idea is to construct the suffix array of the dictionary and use its pattern matching capabilities to parse each document into factors. We now outline the algorithm in more detail.

The suffix array $SA[1..n]$ of a text $x = x[1..n]$ is an array of pointers to all the suffixes of $x$ arranged in lexicographic order, such that, $x[SA[i]..n] < x[SA[i + 1]..n]$. For every substring of $x$, $x[i..j]$, there exists an interval in $SA_x$, such that $SA_x[lb], SA_x[lb + 1], SA_x[...], SA_x[rb]$ contains positions to every occurrence of the substring in $x$. As the suffixes are ordered lexicographically the interval boundaries $lb$ and $rb$ can be calculated with successive binary searches.

The pseudo-code in Figure 1 embodies the RLZ factorization algorithm. Given a string, $x$, and a dictionary, $d$, the function ENCODE computes a factorization of $x$ relative to $d$. This is achieved with successive calls to FACTOR, where we use the suffix array of the dictionary, $SA_d$, for efficient matching. To facilitate document retrieval we stop factorization each time we encounter a document boundary, returning the current $(pos, len)$ pair. We further maintain a document map indicating the offsets of each document boundary in the encoding.

REFINE calibrates the left and right bounds of suffix array, $SA_d[lb..rb]$, such that the suffixes of length $l$ in the interval between these bounds matches the current prefix in $x$, $x[i..i + l]$. The length of the match increases with each successful call to REFINE. As the suffix array is ordered lexicographically, each bound can be calculated using a binary search. An example of this process is demonstrated in Table 1. Here, a dictionary $d = cabbaabba$ of length 9 and its suffix array, $SA_d$, are used to compute a factorization of the input $x = bbaancabb$. We call FACTOR(1, x, d). The first call to REFINE returns the interval $(5, 8)$. The first character of the suffixes between this interval match the first character in $x$. The second call to REFINE returns the interval $(7, 8)$. The suffixes in this interval match the first two characters in $x$. The longest match is 4 characters and occurs in the interval $(8, 8)$. The value of $SA_d[8]$ contains the suffix position in $d$ where the match occurred, in this case 3. The length of the match is 4, so FACTOR will return the pair $(3, 4)$ and we call factor with the new offset FACTOR(5, x, d).

The pseudo-code for the decoding algorithm is described in Figure 2. For each pair, if the length component is 0 then the position contains a character to output. Otherwise, the position value corresponds to an offset in $d$ for which we output the substring $d[position \ldots position + length - 1]$. Decoding is fast as the dictionary is resident in memory.

### 3.3 Dictionary Generation

The dictionary should form a representative sample of the collection. The aim is to capture global repetition across a collection that adaptive compression algorithms do not detect due to their block-oriented nature, or limited window size. We found the following simple approach to be highly effective. We treat a collection as a single string and take evenly spaced samples across the collection. For a collection string, $x = x[1..n]$ of length $n$, we wish to generate a dictionary, $d = d[1..m]$ of length $m$, using samples of length $s$. That is, we take $m/s$ samples at positions $0, n/(m/s), 2n(m/s), \ldots$; in other words, at evenly spaced intervals across the collection. Although simple, we shall see in Section 4 that this technique generates a very effective dictionary for typical Web data.

**Table 1: An example demonstrating the REFINE function. Searching for the string $x$ in dictionary $d$ we refine the range $SA_d[lb..rb]$ such that the longest prefix of $x$ is found in $d$.**

| i | 1 | 2 | 3 | 4 | 5 | 6 | 7 | 8 | 9 |
|---|---|---|---|---|---|---|---|---|---|
| $d[i]$ | c | a | **b** | **b** | **a** | **a** | b | b | a |
| $SA_d$ | 9 | 4 | 8 | 6 | 2 | 3 | 7 | 5 | 1 |

| i | $d[SA_d[i]] \ldots d[len(d)]$ |
|---|---|
| 1 | a |
| 2 | a a b b a |
| 3 | a b b a |
| 4 | a b b a a b b a |
| 5 | **b** a |
| 6 | **b** a a b b a |
| 7 | **b** **b** a |
| 8 | **b** **b** **a** **a** b b a |
| 9 | c a b b a a b b a |

| j | 1 | 2 | 3 | 4 | 5 | 6 | 7 | 8 | 9 |
|---|---|---|---|---|---|---|---|---|---|
| $x[j]$ | **b** | **b** | **a** | **a** | n | c | a | b | b |
| $lb$ | 5 | 7 | 8 | 8 | -1 | | | | |
| $rb$ | 8 | 8 | 8 | 8 | -1 | | | | |

## 3.4 Factor Encoding

Efficient encoding of the $(p, l)$ pairs into which a document is factorized is critical to overall compression effectiveness. We explored several approaches.

Our first approach was to assume the position elements of each pair (the $p$ values) would be spread randomly across the dictionary, and so, difficult to compress. We represented each $p$ value as a single unsigned 32-bit integer, and concentrated on encoding the length elements (the $l$ values). The average factor length recorded from varied combinations of dictionary size and sample length across two document collections remained relatively constant, ranging from 30 to 40 (Table 2 and Table 3). Furthermore, we observed that a significant percentage of length values in an encoding were less than 100, and usually no greater than the sample size used to generate the dictionary. This is illustrated in Figure 3, which plots histograms of encoded length values for a factorization of the GOV2 collection using a 0.5GB dictionary and varied sample periods. Observe, irrespective of the sample period, the bulk of length values remain small. In light of this we used a variable byte (VBYTE) code [26] to encode length values, which provides a reasonable trade-off between compression and decoding speed. Using VBYTE the majority of length values are encoded in just a single byte.

Closer inspection of the position values revealed that while the distribution of the $p$ values across the entire collection was rather flat, applying a compressor (ZLIB) to the $p$ values for each document separately gave a significant boost to compression, suggesting that within each document $p$ values could be quite skewed.

This effect is manifested by substrings that are repeated within a file, but not present in the dictionary. These sub-

**Function** ENCODE($x$, $d$)
   $i \leftarrow 1$
   **while** $i \le len(x)$ **do**
      $(position, length) \leftarrow$ FACTOR($i$, $x$, $d$)
      **output** $(position, length)$
      **if** $length = 0$ **then**
         $i \leftarrow i + 1$
      **else**
         $i \leftarrow i + length$

**Function** FACTOR($i$, $x$, $d$)
   $lb \leftarrow 1$
   $rb \leftarrow len(d)$
   $j \leftarrow i$
   **while** $j \le len(x)$ **do**
      **if** $lb = rb$ **and** $d[SA_d[lb] + j - i] \ne x[j]$ **then**
         – we can no longer refine the interval in $SA_d$ and the current character in $d$ does not match $x[j]$
         **break**
      $(lb, rb) \leftarrow$ REFINE($lb$, $rb$, $j - i$, $x[j]$)
      **if** $(lb, rb)$ is no longer a valid interval **then**
         **break**
      $j \leftarrow j + 1$
      **if** x[j] is at a document boundary **then**
         **break**
   **if** $j = i$ **then**
      **return** $(x[j], 0)$
   **else**
      **return** $(SA_d[lb], j - i)$

**Figure 1: Encoding Algorithm.**

**Function** DECODE($x$, $d$)
   **foreach** $(position, length)$ **pair** $\in x$ **do**
      **if** $length = 0$ **then**
         **output** $position$
      **else**
         **output** $d[position \ldots position + length - 1]$

**Figure 2: Decoding Algorithm.**

strings get factorized into the same set of pairs, which are then repeated in the file's RLZ factorization. Applying a local compressor to the pairs captures these local repetitions and improves overall compression.

We observed the same phenomenon in the length values. That is, they contained higher-order patterns at the document level. We grouped the pairs for each document and compressed the positions and lengths separately using ZLIB. We report on experiments with different combinations of these techniques in Section 4.

## 3.5 URL Sorting

In their study on the compressibility of web pages, Ferragina and Manzini [16] show that arranging the collection of web pages in URL order prior to applying ZLIB (and other standard compressors) leads to a significant improvement in compression performance. For example, ZLIB compresses GOV2 to 18.69% with pages in (natural) crawl order, and to 10.41% in URL order.

While URL sorting can clearly be effective for some web collections, the wider use of this approach seems dubious. In many contexts documents do not have explicit relatedness tags, such as URLs, and so cannot be sorted. Even in web collections, URL ordering can be a poor approximation to real clustering of content, failing to capture, for example, mirrored sites with identical content but different URLs. Also, the technique cannot be easily applied to dynamic document collections, and sorting of a large, disk resident collection is a non-trivial task.

**Table 2: Average factor length and percentage of unused bytes in an RLZ dictionary for varied dictionary and sample sizes built on a 426 GB GOV2 corpus.**

| Size (GB) | Samp. (KB) | Avg.Fact. | Unused (%) |
|---|---|---|---|
| 2.0 | 0.5 | 46.01 | 39.62 |
| 2.0 | 1.0 | 46.83 | 24.28 |
| 2.0 | 2.0 | 46.77 | 28.10 |
| 2.0 | 5.0 | 46.09 | 20.65 |
| 1.0 | 0.5 | 41.30 | 36.00 |
| 1.0 | 1.0 | 41.80 | 31.38 |
| 1.0 | 2.0 | 41.62 | 25.66 |
| 1.0 | 5.0 | 40.93 | 17.84 |
| 0.5 | 0.5 | 37.07 | 32.91 |
| 0.5 | 1.0 | 37.35 | 28.64 |
| 0.5 | 2.0 | 37.15 | 23.65 |
| 0.5 | 5.0 | 36.45 | 16.20 |

**Table 3: Average factor length and percentage of unused bytes in an RLZ dictionary for varied dictionary and sample sizes built on a 256 GB Wikipedia corpus.**

| Size (GB) | Samp. (KB) | Avg.Fact. | Unused (%) |
|---|---|---|---|
| 2.0 | 0.5 | 38.70 | 27.34 |
| 2.0 | 1.0 | 39.11 | 21.33 |
| 2.0 | 2.0 | 39.13 | 17.29 |
| 2.0 | 5.0 | 38.97 | 12.22 |
| 1.0 | 0.5 | 34.54 | 23.72 |
| 1.0 | 1.0 | 34.85 | 18.52 |
| 1.0 | 2.0 | 34.81 | 13.99 |
| 1.0 | 5.0 | 34.63 | 9.56 |
| 0.5 | 0.5 | 31.05 | 21.15 |
| 0.5 | 1.0 | 31.22 | 15.83 |
| 0.5 | 2.0 | 31.17 | 11.53 |
| 0.5 | 5.0 | 30.96 | 7.41 |

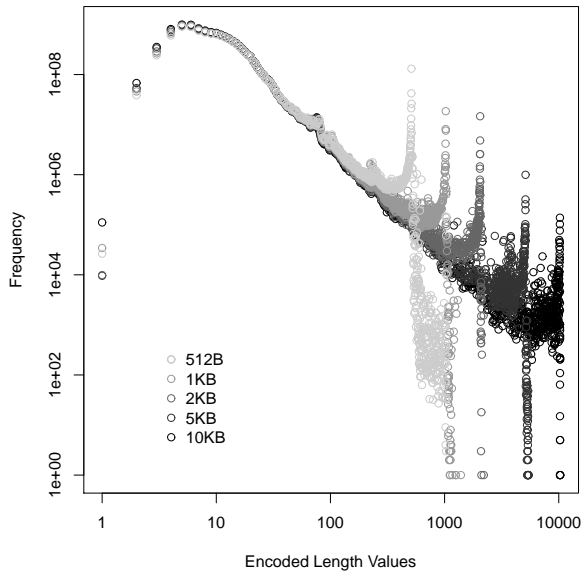

**Figure 3: Frequency histogram of length values in an RLZ encoding of the 426 GB GOV2 corpus using a 0.5 GB dictionary and varied sample periods.**

Because we sample uniformly across the collection rather than build a dictionary based on the order in which documents are encountered, our algorithm mitigates compression effectiveness based on page order.

### 3.6 Dynamic Updates

A further virtue of our method is its application in a dynamic environment where documents are appended to the collection over time. Due to the nature of our sampling process, as long as additional documents maintain similar characteristics to the initial collection there will be little to no impact on compression effectiveness.

If per-document compression degrades below a specific threshold there are several ways to compensate. If there is no constraint on memory, we can sample the new documents and append them to the dictionary. This method avoids an expensive re-encoding process as the previous pair codes are still valid. The suffix array will need to be recomputed in order to include the new samples during factorization. If there are constraints on memory, the dictionary can be regenerated taking the additional documents into consideration. This invalidates the original encoding, and, as a consequence, the collection will need to be compressed again.

## 4. EVALUATION

We evaluate RLZ by comparison against three baseline document retrieval systems.

*Method.* To simulate various aspects of a document retrieval, two access patterns were used throughout our experiments. First, we used a sequential list of 100,000 document IDs to simulate requests for large-scale batch processing. Second, to simulate the typical behavior of a document retrieval system we generated a list of 100,000 document IDs from the ranked output of real queries into a search engine. Each collection was indexed using the Zettair search engine[3], then queried using default settings and a log sourced from topics 20,001 to 60,000 from the 2009 Million Query Track.[4] The top 20 document IDs for each query were concatenated to a list and capped at 100,000.

*Systems Tested.* The first baseline is simply a raw concatenation of uncompressed documents with a map specifying offsets to each document location. We further compare against two standard documents grouped into fixed-size blocks and compressed with an adaptive algorithm. We use ZLIB[5] and LZMA[6] for compression, as these were the two best systems reported in Ferragina and Manzini's extensive study [16]. Block sizes used for both ZLIB and LZMA begin as a single document per block, denoted as 0.0MB, and increase in size to 0.1MB, 0.2MB, 0.5MB and 1.0MB.

---

[3] http://www.seg.rmit.edu.au/zettair
[4] http://trec.nist.gov/data/million.query09.html
[5] http://www.zlib.net
[6] http://www.7-zip.org/sdk.html

RLZ runs are identified by their dictionary size and position–length coding schemes used to compress each document. Methods used were Z, zlib with Z_BEST_COMPRESSION, V, variable byte coding and U, unsigned 32 bit integers. Dictionary sizes used in the evaluation section were 0.5GB, 1.0GB and 2.0GB. Unless stated otherwise, all RLZ dictionaries were generated from 1KB samples.

To evaluate the performance of our method in a dynamic environment we simulate update behaviour by generating dictionaries from fixed prefixes of a collection. We then use these dictionaries to compress the complete collection, observing any impact on compression.

*Test Collections.* Two document collections were used. TREC GOV2 is a 426GB web crawl of the .gov top level domain in 2004. This consists of roughly 25 million documents, with an average document size of 18KB. The second collection is a 256GB English Wikipedia snapshot sourced from Clueweb09,[7] consisting of approximately 6 million documents and an average document size of 45KB. Experiments were conducted on both collections sorted in natural web crawl order. We ran further experiments on the GOV2 collection where documents are sorted by URL.

*Experimental Environment.* All document retrieval experiments were conducted on an Intel Xeon 3.0 GHz processor with 4GB of main memory. The disk was a Seagate Scandisk II, 1Tb, 7200 RPM, with 32 Mb cache. The operating system was Red Hat Enterprise Linux Server release 5.5 (Tikanga), running Linux kernel version 2.6.18. The compiler used was GCC 4.1.2 with full optimizations.

All time results were recorded as wall clock time. As the compressed collections used for evaluation were significantly larger than internal memory it is important to account for disk seek and read latency as they are the dominant cost in document retrieval. We ensured each collection was the only one present on the disk for each run, to eliminate disk position bias. Caches were dropped between each run with `sync && echo 3 > /proc/sys/vm/drop_caches`. We define compression ratio as a percentage of the encoded output against the original collection size.

# 5. DISCUSSION

Compression statistics and document retrieval times for RLZ and block-oriented baselines are shown in Tables 4 to 9.

With the exception of some cases under URL sorting (discussed below), RLZ clearly outperforms ZLIB and LZMA in time and space for both sequential and query-log document request scenarios. Comparing cases with similar memory requirements and compression effectiveness, for sequential access our RLZ approaches a thousand times the speed of the competitor methods. Excepting cases where the compression achieved by the competitor methods is particularly poor, the sequential speed of RLZ is generally at least ten times greater, and the random-access speed is always better by a significant margin.

The effectiveness of RLZ compression validates our dictionary sampling hypothesis that we are capturing global repetitive properties of a collection that the baselines cannot detect. A key factor attributing to the performance of RLZ decoding speed is that the dictionary is static and present in memory. Decoding can start immediately. The compressed baselines incur a penalty initializing and computing a new dictionary for each document request. The baselines are subject to a further penalty having to decode at least half a block on average to retrieve a document.

**Table 4: Sequential and Query-log retrieval speed in documents per second on a 426 GB GOV2 corpus for varied combinations of RLZ dictionaries sizes and position–length codes.**

| Size (GB) | Pos–Len | Enc. (%) | Sequential | Query Log |
|---|---|---|---|---|
| 2.0 | ZZ | 9.26 | 12,857 | 112 |
| 1.0 | ZZ | 9.98 | 10,449 | 113 |
| 0.5 | ZZ | 10.74 | 9,752 | 116 |
| 2.0 | ZV | 9.35 | 18,694 | 110 |
| 1.0 | ZV | 10.17 | 16,591 | 109 |
| 0.5 | ZV | 11.04 | 14,310 | 114 |
| 2.0 | UZ | 10.68 | 15,288 | 109 |
| 1.0 | UZ | 11.87 | 13,902 | 106 |
| 0.5 | UZ | 13.18 | 11,779 | 110 |
| 2.0 | UV | 10.77 | 21,622 | 110 |
| 1.0 | UV | 12.06 | 20,327 | 109 |
| 0.5 | UV | 13.48 | 16,107 | 109 |

**Table 5: Sequential and Query-log retrieval speed in documents per second on a 426 GB URL-sorted GOV2 corpus for varied combinations of RLZ dictionaries sizes and position–length codes.**

| Size (GB) | Pos–Len | Enc. (%) | Sequential | Query Log |
|---|---|---|---|---|
| 2.0 | ZZ | 9.23 | 18,684 | 118 |
| 1.0 | ZZ | 9.97 | 16,994 | 116 |
| 0.5 | ZZ | 10.74 | 15,547 | 114 |
| 2.0 | ZV | 9.32 | 26,461 | 115 |
| 1.0 | ZV | 10.16 | 25,278 | 111 |
| 0.5 | ZV | 11.04 | 22,535 | 99 |
| 2.0 | UZ | 10.72 | 35,189 | 111 |
| 1.0 | UZ | 11.91 | 33,171 | 107 |
| 0.5 | UZ | 13.22 | 29,305 | 97 |
| 2.0 | UV | 10.81 | 80,205 | 113 |
| 1.0 | UV | 12.10 | 73,196 | 108 |
| 0.5 | UV | 13.52 | 48,998 | 97 |

Ordered document requests have much faster decoding rates due to sequential disk access. UV pair coding was the fastest method due to its cheap decoding procedure. ZZ was the slowest method, but it was still much faster than the two baselines and achieves excellent overall compression.

As expected, RLZ sampling on a URL-sorted collection had no impact on compression effectiveness, only varying by a fraction of a percent (refer to Tables 4 and 5). However, there was significant speed increase during sequential document decoding, up to 80,000 documents per second using a 2.0GB dictionary and UZ pair coding. This dramatic increase is due to cache locality of common factors shared between similar documents.

LZMA with a large block is competitive in space with RLZ

**Table 6: Sequential and Query-log retrieval speed in documents per second on a 426 GB GOV2 corpus for baseline ASCII and blocked LZ files.**

| Alg. | Block (MB) | Enc. (%) | Sequential | Query Log |
|------|-----------|----------|-----------|-----------|
| ascii | - | 100.00 | 8,982 | 28 |
| zlib | 0.0 | 24.13 | 6,263 | 96 |
| zlib | 0.1 | 20.54 | 1,509 | 67 |
| zlib | 0.2 | 19.38 | 773 | 53 |
| zlib | 0.5 | 18.66 | 313 | 45 |
| zlib | 1.0 | 18.43 | 153 | 36 |
| lzma | 0.0 | 22.33 | 1,490 | 91 |
| lzma | 0.1 | 17.24 | 338 | 60 |
| lzma | 0.2 | 14.29 | 180 | 47 |
| lzma | 0.5 | 11.92 | 78 | 33 |
| lzma | 1.0 | 10.81 | 41 | 22 |

**Table 7: Sequential and Query-log retrieval speed in documents per second on a 426 GB URL-sorted GOV2 corpus for baseline ASCII and blocked LZ files.**

| Alg. | Block (MB) | Enc. (%) | Sequential | Query Log |
|------|-----------|----------|-----------|-----------|
| ascii | - | 100.00 | 3,312 | 36 |
| zlib | 0.0 | 24.13 | 3,827 | 92 |
| zlib | 0.1 | 13.21 | 2,014 | 78 |
| zlib | 0.2 | 12.03 | 1,076 | 64 |
| zlib | 0.5 | 11.31 | 380 | 48 |
| zlib | 1.0 | 11.08 | 190 | 39 |
| lzma | 0.0 | 22.33 | 899 | 91 |
| lzma | 0.1 | 10.38 | 501 | 81 |
| lzma | 0.2 | 8.12 | 338 | 73 |
| lzma | 0.5 | 6.43 | 172 | 59 |
| lzma | 1.0 | 5.67 | 96 | 43 |

**Table 8: Sequential and Query-log retrieval speed in documents per second on a 256 GB Wikipedia corpus for varied combinations of RLZ dictionaries sizes and position–length codes.**

| Size (GB) | Pos–Len | Enc. (%) | Sequential | Query Log |
|-----------|---------|----------|-----------|-----------|
| 2.0 | ZZ | 9.56 | 7,898 | 125 |
| 1.0 | ZZ | 10.68 | 7,786 | 129 |
| 0.5 | ZZ | 11.77 | 6,932 | 129 |
| 2.0 | ZV | 9.74 | 13,360 | 132 |
| 1.0 | ZV | 10.92 | 12,766 | 130 |
| 0.5 | ZV | 12.07 | 11,156 | 130 |
| 2.0 | UZ | 12.67 | 9,351 | 104 |
| 1.0 | UZ | 14.16 | 9,563 | 105 |
| 0.5 | UZ | 15.74 | 8,557 | 103 |
| 2.0 | UV | 12.85 | 17,422 | 112 |
| 1.0 | UV | 14.40 | 17,979 | 114 |
| 0.5 | UV | 16.05 | 15,834 | 117 |

**Table 9: Sequential and Query-log retrieval speed in documents per second on a 256 GB Wikipedia corpus for baseline ASCII and blocked LZ files.**

| Alg. | Block (MB) | Enc. (%) | Sequential | Query Log |
|------|-----------|----------|-----------|-----------|
| ascii | - | 100.00 | 2,093 | 50 |
| zlib | 0.0 | 24.13 | 2,610 | 98 |
| zlib | 0.1 | 20.54 | 1,690 | 90 |
| zlib | 0.2 | 19.38 | 902 | 80 |
| zlib | 0.5 | 18.66 | 355 | 64 |
| zlib | 1.0 | 18.43 | 172 | 48 |
| lzma | 0.0 | 22.33 | 604 | 93 |
| lzma | 0.1 | 17.24 | 437 | 86 |
| lzma | 0.2 | 14.29 | 271 | 79 |
| lzma | 0.5 | 11.92 | 123 | 55 |
| lzma | 1.0 | 10.81 | 65 | 32 |

methods, and superior when the collection is URL-sorted. In our experiments, lzma with a block size of 1.0MB compressed a URL-sorted GOV2 to 5.67%. However, ordered and query log document request speeds for lzma were the slowest overall: rlz is five times faster on GOV2, when memory is equated and the collection is in natural order, and there is a two-fold improvement when the collection is sorted by URL.

Query log requests were much slower than sequential requests due to latency during disk operations. Focusing on the compressed baselines, as expected, the fastest throughput was achieved by the two implementations where single documents were encoded in each block as there was no additional overhead when decoding a document. At the same time, the single document methods were the largest of the block-oriented encodings as there was less redundancy to exploit. This mirrors results reported elsewhere [16].

A larger dictionary was beneficial for ordered document requests on both collections, but there was no clear benefit to the use of a larger dictionary for query-log document requests. All rlz methods reported consistent access speeds, averaging over 100 documents per second. ZZ and ZV pair coding methods ran slightly faster on the Wikipedia collection. We attribute this to Wikipedia's average document size being much larger, and zlib being able to compress the pairs effectively. Dictionary size and composition is an interesting direction for future work.

Results in Table 10 demonstrate that our algorithm responds well in a dynamic environment, where new documents are added to the collection. In our simulation we generated 1GB dictionaries from 90% to 10% prefixes of Wikipedia. We observed less than 1% difference in compression relative to the original dictionary. Indeed, the loss when using a dictionary from a 1% prefix of Wikipedia was only 1.35%. This shows that rlz should provide a highly robust compression method in the presence of updates.

# 6. CONCLUSION AND FUTURE WORK

We have described rlz, a compression algorithm and dictionary generation technique for large collections that provides both highly effective compression and fast decoding of individual documents. Our algorithm can dramatically outperform state-of-the-art block-oriented techniques, primarily because it is able to capture global repetition, which block-oriented techniques inherently miss. An additional virtue of rlz is its scalability: it is lightweight at compression time, both in principle and in practice.

There are numerous avenues for future work. Although our sampling method successfully captured global duplica-tion and provides excellent compression, there is still redundancy throughout the dictionary, that is, parts of the dictionary are not used. A possible approach is to make multiple passes of random sampling. During each pass we find and eliminate redundancy, freeing space to be filled in subsequent passes. Some preliminary investigations in this direction are reported elsewhere [17]. Another issue is that there are many choices for compression of position and length values with different space-time tradeoffs. While we have shown that higher-order patterns exist in the factor values, and that these can be exploited to improve both space and throughput, alternative integer codes, such as SIMPLE9 [1] or PFORDELTA [36] may substantially improve on VBYTE and give relevant points on the tradeoff curve. However, even without such further developments, RLZ is a practical and efficient method for corpus compression that dramatically outperforms previous techniques.

**Table 10: Compressing a 265 GB Wikipedia corpus using ZZ pair codes relative to a 1 GB dictionary built from varied prefixes of the collection.**

| Prefix % | Encoding % |
|---------|-----------|
| 100.0 | 10.68 |
| 90.0 | 10.70 |
| 80.0 | 10.73 |
| 70.0 | 10.76 |
| 60.0 | 10.89 |
| 50.0 | 11.11 |
| 40.0 | 11.17 |
| 30.0 | 11.25 |
| 20.0 | 11.37 |
| 10.0 | 11.64 |
| 1.0 | 12.04 |

## Acknowledgements

This work was supported by the Australian Research Council and the NICTA Victorian Research Laboratory. NICTA is funded by the Australian Government as represented by the Department of Broadband, Communications, and the Digital Economy, and the Australian Research Council through the ICT Centre of Excellence program. Simon J. Puglisi is supported by a Newton Fellowship.